\documentstyle{article}

\def\QQa{\renewcommand{\baselinestretch}{1.3}\Huge\normalsize\large\small}

\begin{document}
\QQa

\large
\begin{flushright}
ITP.SB-93-14\\

\end{flushright}

\begin{center}
\Large
{\bf Quantum Dynamical Model for Wave Function Reduction in
Classical and Macroscopic Limits\\}
\vspace{1cm}

\Large

Chang-Pu Sun \\

Institute for Theoretical Physics, State University of New York, Stony Brook,
NY 11794-3840, USA\\
and\\
Physics Dpartment, Northeast Normal University,
Changchun 130024, P.R.China\\
\vspace{1cm}
\large
{\bf Abstract}\\
\end{center}
In this papper, a quantum dynamical model describing the quantum measurement
process is presented as an extensive generalization of the Coleman-Hepp model.
In both the classical limit with very
large quantum number and macroscopic limit with very large particle number
in measuring instrument, this model generally realizes the wave packet collapse
in quantum measurement as a consequence of the Schrodinger time evolution
in  either the exactly-solvable case or the non-(exactly-)solvable case.
 For the
latter, its quasi-adiabatic case is explicitly analysed by making use of
the high-order adiabatic approximation method and then
manifests the wave packet collapse as well as the exactly-solvable case.
By highlighting these analysis, it is finally found that an essence of
the dynamical model of wave packet collapse is the factorization of the
Schrodinger evolution other than the exact solvability. So many dynamical
models including the well-known ones before, which are exactly-solvable
or not, can be shown  only to be the concrete realizations of this
factorizability.
\vspace{0.5cm}

---------------------------------------------------------------\\
\vspace{0.2cm}


\large
\newpage

{\bf 1.Introduction}
\vspace{0.4cm}

Though quantum mechanics has been experimently proved as a quite successful
theory, its interpretation is still an important problem that the physicist can
not avoid completely [1-4]. In order to interperate its mathematical
formulism physically, one has to introduce the wave packet collaps(WPC)
postulate  as an extra assumption added
to the  closed system of rules in quantum mechanics. This postulate is
also called von Neumann's projection rule or wave function reduction process.
Let us now describe it briefly. It is well known in quantum physics that, if
measured quatum system S is in a state $|\phi>$ that is
a linear superposition of the
eigenstates $|k>$  of the  operator $\hat A$ of an observable A just before a
 measurement, ie.,
$$|\phi>=\sum c_k|k>,c_k's ~~are~~ complex~~ numbers\eqno{(1.1)}$$
then a result of th measurement of A is one $a_k$ of
the eigenvalues of $\hat A$ corresponding to $|k>$
with the probability $|c_k|^2$. The von Neumann's postulate tell us that, once
a well-determined result $a_k$ about A has been obtained, the state of S is
no longer  $|\phi>$ and it must collapses into $|n>$ since the
immeadiately-successive measurement of A after the first one
should repeats the same result. Using the density matrix
$$\rho =|\phi><\phi|=\sum_{k,k'} c_kc_{k'}^*|k><k'|,\eqno{(1.2)}$$
for the state $|\phi>$, the above WPC process can be  expressed as
a projecton or  reduction
$$\rho\rightarrow \hat {\rho}=\sum |c_n|^2|k><k|.\eqno{(1.3)}$$

However, to realize the WPC, the
external classical measuring apparatus must be used to detect the result.
Then, someone
thinks the WPC postulate to be not quite satisfactory since quantum mechanics
is expected to be an universal theory valid for whole `universe' because the
detector ,as a part of the universe, behaves classicaly in
the von Neumann's postulate.
A reasonable description of the detector should
be quantum essencially and it exhibts the classical or macroscopic
features in certain
limits. If one deal with the detector as a subsystem of the closed system
(universe= the measured system S + the detector D), it is possible
that the quantum
dynamics of the universe can result in the WPC through the interactions
between  S and D. Up to new, some exactly-solvable models have been
 presented to
analyse this problem [5-10]. Among them, the Coleman-Happ (CH)
model is very famous
one and has been extensively studied in last tweenty years [5-9].
In order to describe  studies in this paper clearly, we need to see
some details of this model.

In the original CH model, an  ultrarelativistic particle is
referred to the measured system S while a one-dimensional array of scatterers
with spin-1/2 to the detector D. The  interaction between S and D
is represented
by an homogeous coupling

$$ H_I=\sum_{n=1}^NV( x-a_n)\sigma_1^{(n)}\eqno{(1.4)}$$
where $\sigma_1^{(n)}$ is the first component of Puli matrix; $a_n$ is the
position of the scatterer assigned to the n'th site in the array.
The Hamiltonian for D is
$$H_s=c\hat P\eqno{(1.5)}$$
 where $c$, $\hat P$ and $ x$ are the light speed, the momentum and
 coordinate operators respectively for S.
This model is quite simple, but it can be exactly solved to produce a
deep insight on the dynamical description of the quantum measurement process.
Starting with the initial state
$$|\psi(0)>=\sum c_k|k>\otimes |D>\eqno{(1.6)}$$
where $|D>$ is  pure state of D (it is usually taken to be ground state),
the evalution state $|\psi(t)>$ for the universe =S+D is defined by the
exact solution to this model. Then, the reduced density matrix

$$\rho_s(t)=Tr_D(|\psi(t)><\psi(t)|)\eqno{(1.7)}$$
of the measured system is obtained by taking the trace of the
density matrix
$$\rho(t)=|\psi(t)><\psi(t)|\eqno{(1.8)}$$
 of the universe to the
variables of D. Obviously, $\rho_s(t)$ depends on the particle number N of
D. When $N\rightarrow \infty$, i.e., in the macroscopic limit ,
$\rho_s(t)\rightarrow \hat{\rho}$
 after long enough time t as  eq.(1-2). Namely,
the Schrodinger evolution of the
universe=S+D leads to the WPC for the measured system.
More recently, the original CH model was improved to describe the energy
exchange between S and D by adding a free energy Hamiltonian [9]
$$H_0=\hbar\omega \sum_{n=1}^N\sigma_3^{(n)}\eqno{(1.9)}$$
and correspondingly improving the interaction slightly. Notice that
the improved model remains exctly-solvable.

However,because the spin quantum number is fixed to be 1/2 in the original
CH model or its improved versions, they can not
describe the {\it classical characters } of the measurement. Usually,
the classical
feature of a quantum object is determined by taking certain value for
some internal quntum numbers of the detector D or $\hbar=0$. In the case of
the
angular momentum, this classical limit corresponds to infinite spin.
In a nice paper[10], this problem
was analysed by using another exactly-solvable dynamical model for
quantum measurement. So it is expected that the WPC in classical limit can be
incoporated in an extensive generalization of CH model. The first step
of this paper is to establish such
a generalized CH model manifeasting the WPC as the dynamical
process in the  classical limit as well as in the the macroscopic limit
simultaneously. Then,it is tried to  find the essence  for
this model subtantially
resulting in the realization of the WPC as a quantum dynamical process
as well as for those well-established ones before.
To this end,we will explicitly study the dynamics of this generalized CH model
in both the exactly-solvable case and the non-solvable case. For the latter,
we will apply the high-order adiabatic approximation (HOAA) method
[11-13] to its special case that the coupling parameter depends on
the position of the measured ultrarelativistic particle quite slightly.
Finally,
we pont out the possible essence in the dynamical realization of the WPC,
which is largely independent of the concrete forms of model Hamiltonians.
\vspace{0.4cm}

{\bf 2.Generalization of the CH model and Its Exact Evolution Operator}
\vspace{0.4cm}

Based on the original CH model, the present gneralizations are to assign
an arbitrary spin $j_n$  to each scatterer
on one-dimensional array  as the detector D  and to take an inhomogeneous
coupling of the scatters to the ultrarelativistic particle as the measured
system S . In this case the spin couplings
have different directions  on different sites of the array.
Let
$${\bf J}(n)= (\hat J_x(n),\hat J_y(n), \hat J_z(n))$$
be the angular momentum operator
acting on the n'th site and the angular momentum operators on different sites
$n=1,2,...,N$ commute with each other. Then, we write the interacting
Hamiltonian for the present generalized  model
$$H_I=\sum^{N}_{n=1}{\bf J}(n).{\bf B}(x-a_n),\eqno{(2.1)}$$
in terms of the 3-vectors  ${\bf B}(x-a_n)$  depending on the coordinate
x  of S  and the fixed
coordinates $a_n$ of the scatterers in the spin array. As
the energy-exchanging between D and S is studied in ref.[9], we introduce
a free Hamiltonian for the spin array D
$$H_D=\sum^{N}_{n}B_0(x-a_n)\hat J_z(n),\eqno{(2.2)}$$
to distinguish the states of the detector D via energy levels.
Then, we have a Hamiltonian
$$H=c\hat P + \sum^{N}_{n}{\bf J}(n).{\bf R}(x-a_n),\eqno{(2.3)}$$
for the universe =S+D where
$$ {\bf R}(x)=(B_1(x),B_2(x),B_3(x)+B_0(x)).$$

In the above model, because of the introduction of the {\it arbitrary spin j}
, which labels any 2j+1-dimensional  irreducible representation of rotation
group SO(3), we are able to consider the behaviours of the quatum dynamics
governed by this model Hamiltonian in the classical limit with infinite spin j.
 It will be proved that, like in the macroscopic limit with infinite N,
the quantum dynamical evolution of the universe also lesads to the WPC
for the measured system in the classical limit. The reason that the limit with
infinite j is called  of classical is that the mean squre deviations
of the components $\hat J_x,$ and $\hat J_y $ possess the limit feature[17]
$$\frac{\Delta \hat J_x}{j}=\frac{\Delta\hat J_y}{j}=\frac
{1}{\sqrt{2j}}\rightarrow 0 ,
{}~~as~N \rightarrow 0.$$

To solve the dynamical evolution of the universe=S+D exactly,
 the polar coordinate
$(R,\theta,\phi)$ for the space $\{{\bf R}\}$ of the coupling parameter :
$${\bf R}=R(sin\theta cos \phi,sin\theta cos\phi,cos \theta)$$
is introduced  where
$$R(x)=\sqrt{B_1^2(x)+B_2^2(x)+(B_0(x)+B_3(x))^2},$$
$$tg\theta(x)=\frac{\sqrt{B_1^2(x)+B_2^2(x)}}{B_3(x)+B_0(x)},\eqno{(2.4)}$$
$$tg \phi(x)= \frac{B_2(x)} {B_1(x)}.$$
Notice that the functions $R, \theta,$ and $ \phi$  usally depend on x through
the coupling paramaters {\bf R}.
According to the quamtum rotation theory,the interaction Hamiltonian
$H_I$ can be rewritten as
$$H_I=S(\theta,\phi)^+\sum^N_{n=1}R_n \hat{J}(n)S(\theta,\phi),\eqno{(2.5)}$$
where
$$S(x)=S(\theta(x-a_n),\phi(x-a_n))=
\prod^{N}_{n=1}e^{-i\hat J_z(n)\phi(x-a_n)/\hbar}
e^{-i\hat J_y(n)\theta(x-a_n)/\hbar},\eqno{(2.6)}$$
is a globle rotation of the spin
array generated by the local rotations
$$S_n(x)=e^{-i\hat J_z(n)\phi(x-a_n)/\hbar}
e^{-i\hat J_y(n)\theta(x-a_n)/\hbar}\eqno{(2.7)}$$
 for each  sites. Latter on ,we will
shows that it is just this factorization of the Hamiltonian that leads to
the WPC in quantum measurement through the factorization of the evolution
operator .

For the  evolution operator U(t) of the universe satisfying the
Schrodinger equation with the Hamiltonian (2.3), we introduce  the
`interaction' picture  by

$$U(t)= e^{-ict\hat P/\hbar}U_e(t),\eqno{(2.8)}$$
where $e^{-ict\hat P}$ is the generator for the coherent state as Gaussian
wave packet [17]. In this picture, the reduced evolution operator obeys
an time-dependent Schrodinger equation
$$i\hbar \frac{\partial}{\partial t} U_e(t)=H_e(t)U_e(t),\eqno{(2.9)}$$
with the time-dependent Hamiltonian
$$H_e(t)=\sum^N_{k=1}h_{ek}(t)=\sum^N_{k=1}{\bf J}(n).{\bf R}(x-a_n+ct),
\eqno{(2.10)} $$

Notice that the Schrodinger equation governed by the Hamiltonian H is exactly-
solvable  only for the hamornic case with
$$\theta(x)=constant ~~\theta ,R=constant $$
$$\phi(x)=\frac{\omega x}{c},~~~~~~~~~~~\omega= a~~ real~~ constant
\eqno{(2.11)}$$
To solve the equation (2.9)in this exactly-solvable case, we use the
Rabi-Ramsy-Schwinger's rotating
coordinate techanique . We carry out the transformation on $U_e(t)$

$$U_e(t)=W(t)U_R(t) =\prod^{N}_{n=1}e^{-i\hat J_z(n)
\frac{\omega }{c}(x-a_n+ct)/\hbar}U_R(t),\eqno{(2.12)}$$
Here ,the rotated evolution operator $U_R(t)$ is governed by the rotated
Hamiltonian

$$H_R=W(t)^{-1}H_e(t)W(t)-\omega\sum^N_{k=1}  J_z(n)$$
$$=\sum^N_{n=1}R[ \hat J_x(n)sin \theta+\hat J_z(n)
(cos\theta-\frac{\omega}{R}) ]\eqno{(2.13)}$$
Notice that this is time-independent Hamiltonian.

In terms of
$$\Omega=R\sqrt{1+\frac{\omega^2}{R^2 }-2cos \theta\frac{\omega}{R}}
,\eqno{(2.14)}$$
$$sin \alpha=\frac{R sin \theta}{\Omega},$$
we rewrite the above rotated Hamiltonian as
$$H_R=\sum^N_{n=1}\Omega[ \hat J_x(n)sin \alpha+\hat J_z(n)cos\alpha]$$
$$=\Omega\sum^N_{n=1}e^{-i\hat J_y\alpha/\hbar} \hat J_z(n)
e^{i\hat J_y\alpha/\hbar}.\eqno{(2.15)}$$
{}From the above expression for $H_R$ , the rotated evolution operator $U_R(t)$
follows immediately
$$U_R(t)=e^{-iH_Rt/\hbar}=\prod^N_{n=1}e^{-i\hat J_y\alpha/\hbar} e^{
-i\Omega\hat J_z(n)/\hbar}
e^{i\hat J_y\alpha/\hbar}.\eqno{(2.16)}$$
Therefore,the evalution operator for the  universe
$$U(t)= e^{-ict\hat p/\hbar}\prod^{N}_{n=1}e^{-i\hat J_z(n)
\frac{\omega x}{c}(x+a_n+ct)/\hbar}e^{-i\hat J_y\alpha/\hbar}
e^{-i\Omega\hat J_z(n)/\hbar}e^{i\hat J_y\alpha/\hbar}\eqno{(2.18)}$$
finally is obtained from the the above eqs.(2.8,2.12,2.16).

Here, we should remark that the exact solvability of the above
generalized CH model mainly depends on the harmornic form of the function
${\bf R}(x)$ of x. If it is not hamornic, the above method
 can not work well and  certain approximation methods should be used
to deal with  the evolution operators approximitely in various cases. If the
coupling function
${\bf B}(x)$  depends on x quite slightly and then
the measured ultrrelativistic particle may
move so slowly that the spin states of the scatterer
in the detector can  be excited hardly,
the adiabatic (Born-Opppehimer) approximation or its generealization
can make sense for the problem.  Thereby,  the Berry's geometric phase[14,15]
and the corresponding induced gauge field can be incorportated  in this
dynamical model of the WPC for the quantum mesurement in the adiabatic
cse.
\vspace{0.4cm}

{\bf 3. Dynamical Realization of Wave Packet Collapse :Exactly-Solvable Case}
\vspace{0.4cm}

To consider the dynamical realizabilty of the WPC in the above model for
quantum measuremet, we consider an ideal double-slit interference
experiment. Let a coherent beam of the  ultrarelativistic particles be split
into two branchs represented by the  wave functions $\mid \psi_1>$
and $\mid \psi_2>$ respectively. In the same time , the detector is assigned to
its ground stete
$$\mid 0>=\mid j_1,m_1=-j_1>\otimes\mid  j_2,m_2=-j_2>
\otimes...\mid j_N,m_N=-j_N >,\eqno{(3.1)}$$
where $ |j_k,m_k>~~~(k=1,2,...,N)$ are standard angular momentum states.
The choise of ground state is required by the stable measurement D. Like
the authors in refs.[5-9], we also suppose that  only
the second branch wave $\mid \psi_2>$ interacts with D.
Starting with the
initial state
$$\mid \psi(0)>=(C_1\mid \psi_1>+
C_2\mid \psi_2>)\otimes \mid 0>\eqno{(3.2)}$$
where
$$|C_1|^2+|C_1|^2=1,$$
the evolution operator (2.18) defines the evolution stete at an instant t
in the `interaction ' picture with the  `interaction ' $H_I+H_D$
$$\mid \psi(t)>=C_1\mid \psi_1>\otimes \mid 0>+C_2
\mid \psi_2>])\otimes
U_e(t)\mid 0>,\eqno{(3.3)}$$
Then,we get the  corresponding desity matrix

 $$\rho(t)=|\psi(t)><\psi(t)|=
|C_1|^2|\psi_1(t)><\psi_1(t)\mid\otimes \mid 0><0\mid $$

$$|C_2|^2|\psi_2(t)><\psi_2(t)\mid\otimes U_e(t)\mid 0><0\mid U_e(t)^+
$$
$$  +C_1C_2^*\mid \psi_1(t)><\psi_2(t)\mid\otimes U_e(t)\mid 0><0\mid $$
$$+C_2C_1^*\mid \psi_2(t)><\psi_1(t)\mid)\mid\otimes \mid 0><0\mid U_e(t)^+.
\eqno{(3.5)}$$
In the problem of WPC,
 because we are only intereset in the behaviors of the system S and the
effect of the detector D on it, we only need
the reduced density matrix for S
$$\rho(t)_S=Tr_D \rho(t)=|C_1|^2\mid \psi_1(t)><\psi_1(t)\mid
+|C_2|^2\mid \psi_2(t)><\psi_2(t)\mid+$$
$$(C_1C_2^*\mid \psi_1(t)><\psi_2(t)\mid+C_2C_1^*\mid
\psi_2(t)><\psi_1(t)\mid)<0\mid U_e(t)\mid 0>,\eqno{(3.5)}$$
where $Tr_D$ represents the trace to the variables of the detector .

Obviously, under a certain conditions to be determined , if the vacuum-
vacuum transition amplitude $<0\mid U_e(t)\mid 0>$ vanishs for the detector D,
the coherent terms in eq.(14) vanish  and thus the quantum dynamics
automatically leads to the wave function reduction,

$$\rho(t)_S\rightarrow
\hat{\rho(t)}=|C_1|^2\mid \psi_1(t)><\psi_1(t)\mid+|C_2|^2\mid \psi_2(t)>
<\psi_2(t)\mid.\eqno{(3.6)}$$
Namely, the WPC occurs as quatum dynamical process under these conditions
!

Now, let us prove that these conditions are just the macroscopic limit
and the classical limit,which respectively correspond to the cases with
very large particle number N and very large quantum number $j_n$. To this end,
 we evolute the norm of vacuum-vacuum transition amplitude
$<0\mid U_e(t)\mid 0>$. Using the the explict expression of d-function
$$d^{j}_{m,m'}(\alpha)=<j,m|e^{-i\hat J_y(n)/\hbar}|j,m'>,$$
we have
$$|<0\mid U_e(t)\mid 0>|=|\prod^{N}_{n=1}\sum^{j_n}_{m_n=-j_n}
d^{j_n}_{-j_n,m_n}(\alpha)
d^{j_n}_{m_n,-j_n}(-\alpha)e^{-im_n\Omega t}|$$
$$=|\prod^{N}_{n=1}\sum^{j_n}_{m_n=j_n}\frac{(2j_n)!}{(j_n+m_n)!(j_n-m_n)!}
(cos^2\frac{\alpha}{2})^{j_n- m_n}(sin^2\frac{\alpha}{2})^{j_n+ m_n}
e^{-im_n\Omega t}|$$
$$=\prod^{N}_{n=1}|cos^2\frac{\alpha}{2}e^{-i\Omega t}+sin^2\frac{\alpha}{2}|
^{2j_n},$$
that is,
$$|<0\mid U_e(t)\mid 0>|=|\prod^{N}_{n=1}(1-
sin^2\frac{\Omega t}{2}sin^2\alpha)|^{j_n},\eqno{(3.7)}$$

The above formula is a main result of this paper , which directly
manifests the
WPC in the classical and macroscopic limits. Let us now go into some
details for this conclusion. Notice that in a nontrivial case, $\Omega,
\alpha\neq 0$
and so
 $$|1-sin^2\frac{\Omega t}{2}sin^2\alpha|$$ is usually a positive
number less than 1. Thus, in the classical limit  with $j_n \rightarrow \infty$
mentioned before,
$$
|<0\mid U_e(t)\mid 0>|\rightarrow 0,as~~j_n \rightarrow \infty.$$
This means $<0\mid U_e(t)\mid 0>\rightarrow 0,~as~~j_n \rightarrow \infty$,
that is to say, the WPC occurs as a quantum dynamical process in
classical limit! This is just
what we expected. Then, we reach  a concise statement that {\it if the detector
behaves classically,but need not to behave macroscopically,the WPC
can be dynamically realized in the measurement.} The classical detector was
required as a purely classical object before, but here it is proved to be
a classical limit of a quantum object and the quantum mechanics can work
well on it for quantum measurement. We should also stress on that the
macroscopic limit with very large N is not necessary for the WPC. So long
as the detector is in classical limit, the WPC still appears as a dynamical
evolution even for  small N.

Now, we turn to discuss  the macroscopic limit behaviours of the problem
in details. In eq.(3.3), let us defined the positive number $\Delta_n(t)$ by

$$e^{-\Delta_n(t)}= [1-sin^2\alpha
sin^2\frac{\Omega t}{2})]^{j_n}\leq 1,\eqno{(3.8)}$$
Then,

$$|<0|U_e(t)|0>|= exp[-\sum^N_{n=1}\Delta_n(t)],\eqno{(3.9)}$$
 Usually, $\Delta_n(t)$ is a non-zero  and positive and thus the series
$\sum^{\infty}_{n=1}\Delta_n(t)$ diverges to infinity, that is to say,
 $ <0|U_e(t)|0>$ as well as its norm   approach zero as $N\rightarrow \infty$.
This just shows
that the WPC can be realized as a quantum dynamical process for the generalized
CH model in the macroscopic limit .
\vspace{0.4cm}

{\bf 4.Adiabatic Approximation for Non-solvable Case}
\vspace{0.4cm}

As  most of the previous studies about the dyanamical realization of
the WPC for quantum measurement, the above discussions in this paper only
concern an extremely idealized case that the model is exactly-solvable.
So it seems  that the exactly-solvability is necessary for this problem.
However, it is not true really. We will observe that the WPC can also happen
in the non-solvable case of the above generalized CH model. In such case,
 the parameter {\bf R}(t) is not harmonic and so
some approximation methods are needed
to probe the evolution of the universe =S+D. As an example of  non-solvable
model, the adiabatic case that
the parameter $R(x+ct-a_n)$ in  $H_e(t)$ depends
on time `slightly' will be used to illustate the above-mentioned
onservation. Because the quasi-energy state of $H_e(t)$ can hardly
be excited by the variation of $H_e(t)$ as t in this case,
the so-called high-order adiabatic approximation (HOAA) method in connection
with Berry's geometric phase [14,15] can effectively be employed to this end.
This method was recently developed  by this
author [11-13] and now reformulated in the evolution operator form
in the appendix.This refoemulation of the HOAA method is quite convenience
for the appplication in this paper.

Defining the functions
$$f_n(t)=f(x-a_n+ct)$$
for $f=R,\theta,\phi$, etc.,
we first factorize the effective evolution operator $U_e(t)$ into
$$U_e(t)=S(t)U'(t)=\prod^{N}_{n=1}e^{-i\hat J_z(n)\phi_n(t)/\hbar}
e^{-i\hat J_y(n)\theta_n(t)/\hbar}U'(t)\eqno{(4.1)}$$
according to the HOAA method. Then, in the equivelent Hamiltonian goverening
$U'(t)$
$$H'(t)=H_0(t)+V(t):$$
$$H_0=
\sum^{N}_{n}[R_n(t)-cos\theta_n(t) \frac{\partial}{\partial t}
\phi_n(t)] J_z(n),\eqno{(4.2)}$$
$$V(t)=\sum^{N}_{n}[-\frac{\partial}{\partial t}\theta_n(t) J_y(n)
+sin\theta_n(t) \frac{\partial}{\partial t}\phi_n(t)
\hat J_x(n)]\eqno{(4.3)}$$
can be regarded as pertubation. The standard pertubation theory
detemines the first-approximate evolution operator
$$U'_0(t)=\prod^{N}_{n=1}e^{-i\int_0^tR_n(t') dt'/\hbar}
e^{-i\int_0^t cos\theta_n(t')\frac{\partial}{\partial t'}\phi_n(t')dt'
\hat J_z(n)}$$
$$=\prod^{N}_{n=1}e^{-i\int_0^tR_n(t') dt'}e^{i\gamma_n(t)J_z(n)},
\eqno{(4.4)}$$
which describes the geometric feature of the evolution in terms of the
Berry's phase
$$\gamma_n(t)=-\int_0^t\frac{\partial}{\partial t'}\phi_n(t')
cos \theta_n(t')dt',\eqno{(4.5)}$$
When the parameter {\bf R} is subject to a cyclic evolution that
{\bf R}(0)={\bf R}(T), the Berry's phase
$$\gamma_n(T)=\int_0^{2\pi}[1-
cos \theta_n]d\phi_n,\eqno{(4.6)}$$
is just a solid angle spanned by the closed corve traced by the
parameter {\bf R}. To consider whether the WPC happen or not for the adiabatic
evolution, we explicitly calculate
$$
|<0|U_e(t)|0>|=\prod^{N}_{n=1}|<0|e^{-i\hat J_y(n)\theta_n(t)/\hbar}
|0>|$$
$$=\prod^{N}_{n=1}|d^{j_n}_{-j_n,-j_n}(\theta_n(t)|^{2j_n}
=\prod^{N}_{n=1}|cos(\theta _n(t)/2)|^{2j_n}.\eqno{(4.7)}$$
By the proof similar to that  in last section,we see that $|<0|U_e(t)|0>|
\rightarrow 0 $ as $N\rightarrow \infty$.Namely, even in a non-solvable case
,the generalized CH model sill realize the WPC quantum dynamically for
the adiabatic evolution.

Furthermore, let us prove that it does so  for the
non-adiabatic evolution. In fact,if the the parameter {\bf R} does not
changes slowly eneough, the adiabatic condition
$$|\hbar\frac{\partial}{\partial t}\phi_n(t)/R_n(t)|,
  |\hbar\frac{\partial}{\partial t}\theta_n(t)/R_n(t)| \ll 1,\eqno{(4.8)}$$
we ,at least ,consider the second order approximation
$$U'_e(t)=U'_0(t)[1+U'_1(t)]=U'_0(t)\prod^{N}_{n=1}[1+U'^n_{1}(t)]
$$
$$= \prod^{N}_{n=1}e^{-i\int_0^tR_n(t')\hat J_z(n) dt'/\hbar}e^{-
i\int_0^t cos \theta_n(t')\frac{\partial}{\partial t'}
\phi_n(t')dt'\hat J_z(n)}$$
$$\{1+\frac{1}{i\hbar}\int_0^tU'_0(t)^\dagger
[-\frac{\partial}{\partial t'}\theta_n(t')\hat J_y(n)
+ sin\theta_n(t')\frac{\partial}{\partial t'}\phi_n(t')\hat
J_x(n)]U'_0(t)dt'\},
\eqno{(4.9)}$$
Because of the cut-off in the Dyson series for the approximate
evoulution operator,
the unitarity of evoulution operator is broken and so its leaded
evolution state is not normalized to unity. Thus, when we calculate
the vacuum-vacuum transition amplitude  $<0|U'_e(t)|0>$ , we should first
renomalized it . Let us by $\tilde{U}'_e(t)$ denote the  renormalized evolution
operator defined by
$$\tilde{U}'_e(t)|\phi>=\frac{U'_e(t)|\phi>}{<\phi| U'_e(t)^\dagger
U'_e(t)|\phi>}.\eqno{(4.10)}$$
for any state vector $|\phi>$. This  renormalization  results in reasonable
vacuum-vacuum transition amplitude satisfying
$$|<0|\tilde{U}'_e(t)|0>|=|\frac{<0|U'_e(t)|\phi>}{<0| U'_e(t)^\dagger
U'_e(t)|0>}|$$
$$=|\prod^{N}_{n}\frac{<j_n,-j_n| U'^n_{1}(t)|j_n,-j_n>}
{<j_n,-j_n| U'^n_{1}(t)^\dagger
U'^n_{1}(t)|j_n,-j_n|>}|.\eqno{(4.11)}$$
As the formula given by eq.(4.11), the above equation also explicitly
defines the  dynamical realization of the WPC in the classical limit
with $N\rightarrow \infty$. Here, we have taken it into account that
$$|\frac{<j_n,-j_n| U'^n_{1}(t)|j_n,-j_n>}{<j_n,-j_n| U'^n_{1}(t)^\dagger
U'^n_{1}(t)|j_n,-j_n>}|\leq 1,for~~ n=1,2,...N.\eqno{(4.12)}$$

Based on the above discussions on the first- and second -order
approximations, we guess that the WPC can be realized in arbitrary order
approximation. Trying to prove  this guess,we find some essential
properties related to the WPC closely in next section.
\vspace{0.5cm}

{\bf 5.  Comments on Essence of Dynamical realizability }
\vspace{0.4cm}

Including the above discussion in this paper, the previous investigatons on the
dynamical realization of the WPC in terms of quantum dynamical models only
proceeded with the concrete form of the model Hamiltonians, especially
of the interactions betwen S and D. It seems that the  dyanamical realizablity
of WPC depends on the choice of concrete forms of interaction
. However, motivated by the
above discussions, we will shows a more universal fact that it is the
factorizability of the evolution, other than its  exactly-
solvability, that leads to the WPC in quantum measurement.
Now, let us describe what
is  the factorization of the evalution. Let x and p be the  cooordinate and
momentum  operator of the measured system respectively; $x_n$(n=1,2,...,N)
 be the variables for the measuring instrument. Usually, the evolution
operator
$U(t,p,x,x_i)$
for the universe =S+D depend on x, p and  $x_n$(n=1,2,...,N). If this operator
can expressed as the following factorizable form
$$U(p,x,x_i)=U_s(p,x,t)\prod^{N}_{n=1}U^{[n]}(x,x_n,t) ,\eqno{(5-1)}$$
then we say that the evolution characterized by  $U(t,p,x,x_i)$ is
 factorizable.
 Here,  $U_s(p,x,t)$ is the evolution operator of D in absence
of the interaction
with  the detector D and the unitary operator $U^{[n]} (x,x_n,t)$
only depend on $x_n$ and $x$ for fixed n. In this case, the reduced
density matrix of
S for the above mentioned double- slit interference  experiment
in the `interaction' picture is obtained as
$$\rho(t)_S=Tr_D \rho(t)=|C_1|^2\mid \psi_1(t)><\psi_1(t)\mid
+|C_2|^2\mid \psi_2(t)><\psi_2(t)\mid+$$
$$[C_1C_2^*\mid \psi_1(t)><\psi_2(t)\mid+C_2C_1^*\mid \psi_2(t)><\psi_1(t)\mid)
]<0\mid U_e(t)\mid 0>.\eqno{(5-2)}$$
where
$$\mid 0>=\mid 0_1>\otimes\mid 0_2>\otimes...\mid 0_N>,$$
 $\mid 0_k>$ is the ground state of  each single particles in D.

Because
$$|<0_k\mid U^{[k]}(t)\mid 0_k>|
=[1-\sum_{n\neq 0}\mid <n\mid U^{[k]}\mid 0_k>|^2]^
{1/2} \leq 1,\eqno{(5-3)}$$
for the possitive function
$$
\Delta_k(t)=-ln(|<0_k\mid U^{[k]}\mid 0_k>|),$$
the series $\sum^{\infty}_{k=1}\Delta_k(t)$ diverges to infinity.
That is to say,
 $ <0|U_e(t)|0>$ as well as its norm
$$|<0|U_e(t)|0>|=\prod^{N}_{k=1}|<0_k\mid U^{[k]}\mid 0_k>|
= exp[-\sum^N_{k=1}\Delta_k(t)],\eqno{(5-4)}$$
approach zero as $N\rightarrow \infty$. Then, the WPC appears in the
macroscopic limit. If we can incoporate a quantum number $J_n$ into
 $U^{[n]}(x,x_n,t)$ such that $\Delta_k(t)\rightarrow \infty$ as $N
\rightarrow
\infty$. When $J_n$ enjoys the classical limit at $J_n=\infty$, like the spins
$j_n$ in
this paper, the WPC also occurs in this limit as a quntum dynamical process.
 Therefore, we conclude  that the essence of the dynamical realizability
of WPC is the factorization of the evolution operator for the appreciated
model of quantum measurement.

Naturally, the succeeded question is what is the general form of the model
Hamiltonian which can realize this factorizable evolution. The answer
is that the Hamiltonian should be decomposable in certain sense. The following
Hamiltonian sufficiently enjoys the answer
$$H=H_0+H'=H_0+H_I +H_D:$$
$$H_I=\sum^N_{k=1}V_k(x,x_k),H_D=\sum^N_{k=1}h_k(x_k),\eqno{(5-5)}$$
Here, the measurd system S is sill represented by an
ultrarelativisic particle with the free Hamiltonian $H_0=c\hat P$, but
the detector D is made of N particles with quite general
single-particle Hamiltonian
 $h_k(x_k),(k=1,2,...,N)$ which is Hermitian. S is assumed to be {\it
independetly} subjected to the interaction $V_k(x,x_k)$ of each particle k.
Here,x and $x_k$ are the coordinates of S and the single particle
k in D respectively and the k'th interaction potential $V_k(x,x_k)$
only depeds on $x$ and
$x_k$ and  $h_k(x_k)$ on the single particle variable  $x_k$
. To prove the above statement, we take the transformation
(2.8). Then
the reduced evalution operator $U_e(t)$ obeys an efficvive
Schrodinger equation with the effective Hamiltonian
$$H_e(t)=\sum^N_{k=1}h_{ek}(t)=\sum^N_{k=1}[h_k(x_k)+V_k(x+ct,x_k)],\eqno{(5-6
)}
$$
depending on time. Since $H_e(t)$ is a direct sum of the time-dependent
Hamiltonians  $h_{ek}(t)$
(k=1,2,...,N) parameterized by $x$, the x-depenedent evolution operator, as
the formal solution to the effictive Schrodinger equation,
$$U_e(t)=\prod^N_{k=1}\otimes U^{[k]}(t)
=U^{[1]}(t)\otimes U^{[2]}(t)\otimes ....\otimes U^{[N]}(t),\eqno{(5-7)}$$
is factorizable, that is to say, $U_e(t)$ is a direct product of the
single-particle
evalution operators
$$U^{[k]}(t)=\Im exp[(1/i\hbar \int_{0}^{t}h_{ek}(t)dt],\eqno{(5-8)}$$
where $\Im$ denotes the time-order operation. As proved in as follows,
it is just the above factorizable property of the reduced evolution
operator that results in the quantum dynamical realization of the WPC. Notice
that some results of  this section was announced by this author more recently
[16].

Before concluding this paper, we shall give
some remarks on the results and method
of this paper. We first point out that this paper emphasizes the unified
desciption of classical limit and macroscopic limit for quantum mesurement.
Because the macroscopic phenomenona in quantum mechanics can not be identified
with those classical ones completely (e.g, the magnetic flux quantization
is a macroscopic quantum phenomenon, but it is not definitely classical), it
is quite necesarry to distinguish between these two cases.
We should also remark  that ,in practical problems,
there must exist interactions among the particles constituting the detector
D, but in  the present discussions  there
is not interactions among the particles in the detector. We understand it
as an ideal case. How
to realize the quantum measurement both for the WPC in the interaction case
is an open question we must face. It is expected , at least for some
special case , that the certain canonical (or unitary)transformation
possibly enable
these particles to become the quasi-free ones. This is just similar to
 the the system
of harmonic osscilators with quadric coupling. In this case,
we can imagine that
the detector is made of free quasi-particles that do not interact with each
other.

\newpage

{\bf Appendix:\\
Reformulation of the High-Order addiabatic Approximation Method}
\vspace{0.5cm}

In order to use it in this paper conveniently, we now reformulate the
the high-order adiabatic approximation (HOAA) method in refs.[11-13]
in a general form , which can work well on the evolution operator
for both the Hermitian  and non-Hermitian Schrodinger time evolutions.

Let the Hamiltonian $H_e(t)$ of the quantum system  depend on time t
through a set of the slowly-changing parameters
${\bf R}(t)=(R_1(t),R_2(t),...,R_K(t))$.
We also assume the quasi-energy levels $E_k(t)$ (k=1,2,...K)
 of the time-dependent Hamiltonian $H(t)=H[R(t)]$ for a frozen time t
are not degenerate. Diagonalize H(t) by a similarity transformation
$S(t)=S[R(t)]$ in the following way
$$S(t)H_e(t)S(t)^{-1}=H_d(t)=\left[\begin{array}{lccr}
E_1(t) & 0      & ..... & 0 \\
0      & E_2(t) & ..... & 0 \\
...    &...     & ..... & ... \\
0      &     0  & ..... & E_K(t)
\end{array} \right ]. \eqno{(A-1)}$$
The correponding quasi-energy state to $E_k(t)$ (k=1,2...,k)is
denoted by $|k(t)>$.

If we  detemine  a solution of the Schrodinger equation of evolution
operator $U_e(t)$ governed by  $H_e(t)$ as the following form
$$U_e(t)=S(t)U'(t), \eqno{(A-2)}$$
then $U'(t)$ obeys the Schrodinger-type equation
$$i\hbar \frac{\partial}{\partial t}U'(t)=H'(t)U'(t), \eqno{(A-3)}$$
where the the equivalent Hamiltonian
$$H'(t)=H_d(t) -i\hbar S(t)^{-1} \frac{\partial}{\partial t}S(t) \eqno{(A-4)}$$
can be decomposed into the diagonal part
$$H_0(t)=H_d(t)+ diagonal~~part~~of ~~[ -i\hbar S(t)^{-1}
 \frac{\partial}{\partial t}S(t) ]  \eqno{(A-5)}$$
and the off-diagonal part
$$V(t)= ~~off-diagonal~~part~~of~~[ -i\hbar S(t)^{-1}
 \frac{\partial}{\partial t}S(t) ] . \eqno{(A-6)}$$
Physically, since V(t) completely vanishes when H(t) is independent
of time , we deduce that V(t) is a pertubation in  the case that
H(t) depends time quite `slightly'. Leter on , we will give the analystic
condition that  V(t) can regarded as a pertubation. Then, the adiabatic
Dyson series  solution  of $U'(t)$
$$U'(t)=U'_0(t)[1+\sum^{\infty}_{k=1}U'_k(t)]:$$
$$U'_0(t)=e^{ \frac{-i}{\hbar}\int ^t_0 H_0(s)ds} \eqno{(A-7)}$$
$$ U'_k(t)=(\frac{-i}{\hbar})^k \int^{t}_0 \int^{s_1}_0 \int^{s_2}_0...
 \int^{s_{k-1}} \bar{V}(s_1)\bar{V}(s_2)...\bar{V}(s_{k-1})\bar{V}(s_k)ds_1
ds_2...ds_{k-1}ds_k$$
immeadiately follows from the standard time-dependent pertubation theory.
Here,
$$\bar{V}(t)=e^{ \frac{i}{\hbar}\int ^t_0 H_(s)ds}V(t)
e^{ \frac{-i}{\hbar}\int ^t_0 H_(s)ds} \eqno{(A-8)}.$$

The first order approximate solution $U'_0(t)$
can be decomposed into the dynamical factor

$$\left[\begin{array}{lccr}
e^{-i\int^t_0dt'E_1(t')/\hbar} & 0      & ..... & 0 \\
0      &e^{-i\int^t_0dt' E_2(t')/\hbar} & ..... & 0 \\
...    &...     & ..... & ... \\
0      &     0  & ..... &e^{-i\int^t_0dt' E_K(t')/\hbar}
\end{array} \right ] \eqno{(A-9)}$$
and the geometric factor
$$\left[\begin{array}{lccr}
e^{i\int^t_0dt'A_1(t')} & 0      & ..... & 0 \\
0      &e^{i\int^t_0dt' A_2(t')} & ..... & 0 \\
...    &...     & ..... & ... \\
0      &     0  & ..... &e^{i\int^t_0dt' A_K(t')}
\end{array} \right ] \eqno{(A-10)}$$
where
$$A_n(t)=i<n|S(t)^{-1}\frac{\partial}{\partial t}S(t)|n> \eqno{(A-11)}$$
and each diagonal element in the above matrix is just the Berry's
phase factor
$$\gamma_n(t)=\int ^t_0A_n(s)ds=i<n|\int ^t_0S(s)^{-1}
\frac{\partial}{\partial s}S(s)
ds|n>.\eqno{(A-12)}$$
In terms of the concept of differential manifold, this phase can be rewriteen
as a curve integral
$$\gamma_n(t)=\gamma_n[{\bf R}(t)]=\int _{{\bf R} (t)}A_{\mu}[{\bf R}]
dR^{\mu} \eqno{(A-13)}$$
of the one-form $A_{\mu}[{\bf R}]dR^{\mu}:$
$$A_{\mu}[{\bf R}]=i<n|S[{\bf R}]^{-1}\frac{\partial}{\partial R^{\mu}}
S[{\bf R}]|n> \eqno{(A-14)}$$
on the parameter manifold
$${\bf M}=\{ {\bf R}=(R_1,R_2,...,R_K)|R_i \in real~number~field,i=1,2,...K\}.
$$
In this sense, the Berry's phase factor $e^{i\gamma_n[{\bf R}(T)]}$ can be
understood as an element of the holonomy group for a closed parameter curve
$C:\{{\bf R}(t)|{\bf R}(0)={\bf R}(T)\}$.

Because the transformation $S[{\bf R}]$ diagonalizing $H_e(t)$ is
not unique,i.e., $S'[{\bf R}]=S[{\bf R}]X[R]$
also diagonalizes  $H_e(t)$ if matrix $X[R]$ commuts with $H_e(t)$. This
means that state vectors $\bar { |n[{\bf R}]>}= (S[{\bf R}]X[R])^{-1}|n>$
as well as $ |n[{\bf R}]>= (S[{\bf R}])^{-1}|n>$
are the instanteneous eigenfunctions of $H_e[R(t)]$. The above indeterminecy
of $S[{\bf R}]$ results in the gauge transformation
$$A_{\mu}[{\bf R}]\rightarrow A'_{\mu}[{\bf R}]= A_{\mu}[{\bf R}]
+i<n|X[{\bf R}]^{-1}\frac{\partial}{\partial R^{\mu}}
X[{\bf R}]|n> .\eqno{(A-15)}$$

{}From the second order approximation
$$U(t)' \simeq U'_0(1)[1+U(t)_1']=e^{ \frac{-i}{\hbar}\int ^t_0 H_0(s)ds}
[1-\frac{i}{\hbar} \int^{t}_0\bar {V} (s)ds],\eqno{(A-16)}$$
we observe that the adiabatic condition,under which the adiabatic approximation
solution $U'_0(t)$ work well,is
$$|\frac{\hbar<n|V(t)|m>}{E_n(t)-E_m(t)}|\ll 1.\eqno{(A-17)}$$

\vspace{0.5cm}

{\bf Acknowledgements\\}

The author wishs to express his sincere thanks to Prof.C.N.Yang for drawing h
is
attentions to new progress in the quantum measurement. He also thank D.H.
Feng,L.H.Yu and W.M.Zhang for many useful discussions.
 He is supported by Cha Chi Ming fellowship through the CEEC in State
University of New York at Stony Brook and in part by the
NFS of China through Northeast Normal University.

\newpage

{\bf References}
\begin{enumerate}
\item J.von Neumann,{\it Mathematische Grundlage de Quantummechanik,}
Springer, Berlin,1933.
\item V.B.Braginsky,F.Y.Khalili,{\it Quantum Measurement},Cambridge,1992.
\item R.Omnes,Rev.Mod.Phys.64(1992),339.
\item M.Gell-mann,J.B.Hartle,{\it Classical Equation of Quantum system},
to be published in Phys.Rev.D.
\item K.Hepp,Helv.Phys.Acta.45(1972),237.
\item J.S.Bell,Helv.Phys.Acta.48(1975),93.
\item T.Kobayashi,Found.Phys.Lett.5(1992),265.
\item M.Namik,S.Pascazio,Phys.Rev.A44.(1991),39.
\item H.Nakazato,S.Pascazio,Phys.Rev.Lett.70(1993) 1.
\item M.Cini,Nuovo Cimento,73B(1983),27.
\item C.P.Sun,J.Phys.A,21(1988)1595.
\item C.P.Sun ,Phys.Rev.D.38(1988),2908.
\item C.P.Sun,Phys.Rev.D,41(1990),1318.
\item M.V.Berry,Proc.R.Soc.Lond.A,392(1984),45.
\item A.Shapere,F.Wilczek,{\it Geometrical Phases in Physics,} World Scientific
 ,Singapore,1990.
\item C.P.Sun,{\it  Dynamical Realizability for Quantum Measurement and
Factorization
of Evolution Operator},preprint:ITP.SB-93-13,1993
\item W.M.Zhang,D.H.Feng,R.Gilmore,Rev.Mod.Phys.62(1990) ,867

\end{enumerate}

\end{document}